\documentclass[sigconf]{acmart}

\AtBeginDocument{%
  }

\setcopyright{acmcopyright}
\copyrightyear{2018}
\acmYear{2018}
\acmDOI{XXXXXXX.XXXXXXX}

\acmConference[Conference acronym 'XX]{Make sure to enter the correct
  conference title from your rights confirmation emai}{June 03--05,
  2018}{Woodstock, NY}
\acmPrice{15.00}
\acmISBN{978-1-4503-XXXX-X/18/06}

\usepackage{multirow}
\usepackage{caption}
\usepackage{subcaption}
\usepackage{graphicx}
\usepackage[ruled,linesnumbered]{algorithm2e}
\usepackage{amsmath,environ}

\begin{document}

\title{Deep Multi-Representation Model for Click-Through Rate Prediction}

\author{Shereen Elsayed}
\email{elsayed@ismll.uni-hildesheim.de}
\affiliation{%
 \institution{Information Systems and Machine Learning Lab}
  \city{Hildesheim}
  \country{Germany}
}

\author{Lars Schmidt-Thieme}
\email{schmidt-thieme@ismll.unihildesheim.de}
\affiliation{%
  \institution{Information Systems and Machine Learning Lab}
  \city{Hildesheim}
  \country{Germany}
}




\begin{abstract}
  Click-Through Rate prediction (CTR) is a crucial task in recommender systems, and it gained considerable attention in the past few years. 
The primary purpose of recent research emphasizes obtaining meaningful and powerful representations through mining low and high feature interactions using various components such as Deep Neural Networks (DNN), CrossNets, or transformer blocks. In this work, we propose the Deep Multi-Representation model (DeepMR) that jointly trains a mixture of two powerful feature representation learning components, namely DNNs and multi-head self-attentions. Furthermore, DeepMR integrates the novel residual with zero initialization (\textbf{ReZero}) connections to the DNN and the multi-head self-attention components for learning superior input representations. Experiments on three real-world datasets show that the proposed model significantly outperforms all state-of-the-art models in the task of click-through rate prediction.

\end{abstract}


\maketitle

\section{Introduction}

On a daily basis, millions of users interact with different websites to search for various products, which is reflected in the ever-increasing importance of online advertisements and recommender systems. Consequently, accurately predicting the click-through rates (CTR) plays an essential role in building effective personalized recommender systems that can positively impact online businesses' revenue. As a result, several machine learning approaches have been proposed over the last decade to focus on improving the performance of the CTR prediction in various recommendation settings.

One of the main aspects of CTR prediction is effectively learning-rich latent representations of the different input fields. This aspect has been tackled by applying various techniques, beginning with wide and deep networks \cite{cheng2016wide}, factorization machines-based models \cite{he2017neural,guo2017deepfm,lian2018xdeepfm,yu2019input,lu2020dual} which rely on inner-product between the latent representation of the input fields and hence capturing low-order feature interaction. On the other hand, other approaches rely on deep neural networks (DNNs), and cross networks \cite{wang2021dcn,wang2017deep}  that can learn feature embedding in a very effective manner. However,  both DNNs and cross networks fail to capture the multiplicative relations across the input fields,  limiting the expressiveness of the resulting latent representations, leading to suboptimal predictive performance.
Recently, transformer-based models have gotten much attention by showing remarkable performances as they successfully capture the multiplicative (high-order) relations by applying the multi-head self-attention mechanism \cite{song2019autoint,lu2020dual}. Nevertheless, there is still a massive untapped potential in combining the latent representation and strengths of the different approaches for achieving superior prediction performance. 

In this paper, we propose a novel mixture of experts model DeepMR to capture effective feature interactions that combine the benefits of DNN-based models and attention-based models via a parallel architecture. 

Noting the cross networks' success in various recent works highlights the importance of integrating residual connections within the deep learning models. Moreover, recent residual techniques such as \textbf{ReZero} \cite{bachlechner2021rezero} have shown remarkable performance by employing learnable weights to the residual connections that tend to better signal propagation in the model and faster convergence and thus obtaining more expressive representations.

To leverage these benefits, our proposed DeepMR model integrates ReZero connections in its DNN and multi-head self-attention blocks allowing it to achieve superior prediction performance compared to other state-of-the-art models. The main contributions can be summarized as follows:
\begin{itemize}
\item We propose DeepMR, a novel mixture of experts model for CTR prediction that utilizes a DNN and multi-head self-attention component with ReZero connections.
\item We evaluate the proposed model on  three real-world publicly available datasets for CTR prediction. Results show that the proposed DeepMR model significantly outperforms all state-of-the-art models in CTR prediction in Area Under the ROC Curve (AUC) and Logloss. 
\item We conduct a comprehensive ablation study to illustrate the importance of each component in the model.  
\end{itemize}


\section{Related Work}
There has been a plethora of related work that focuses on improving the click-through rate prediction performance throughout the last few years. The main aim of the different works is to learn informative fields representation for the input interactions. A famous early work presented by Google is the Deep and Wide model \cite{cheng2016wide}, which benefits from combining the wide component, which helps to memorize the feature interactions while including the DNN component, which maintains the model generalization.\\
\textbf{Factorization Machines and CrossNets based models.} In light of understanding more informative feature interactions, Factorization Machines (FM) \cite{rendle2010factorization} have been broadly used, which enable the model to capture the second-order feature interactions and works effectively with sparse data. Neural factorization machines (NFM) \cite{he2017neural} adopt FM for capturing the second-order interactions and benefit from learning non-linear feature interactions by employing DNNs. Moreover, factorization machine based neural network (DeepFM)\cite{guo2017deepfm} utilized the FM component and DNN to learn low-high feature interactions and obtain effective field representations. An improved version of the DeepFM model named extreme deep factorization machine (xDeepFM) \cite{lian2018xdeepfm} which combines functionalities from convolutional neural networks (CNN) and recurrent neural networks (RNN). Instead, xDeepFM proposes the compressed interaction network (CIN) to learn degree bounded feature interactions as the degree increases with increasing the depth of the network. Furthermore, input aware factorization machines (IFM) \cite{yu2019input} use a factor estimating net to learn input-aware factors for the same input feature across different instances. In this regard, another model, dual input aware factorization machines (DIFM) \cite{lu2020dual} improved over the IFM model by utilizing multi-head self-attention and DNN simultaneously to reweight the feature representations. Other models \cite{wang2021dcn,wang2017deep} rely on Cross Networks which produce the feature crosses to learn bounded degree feature interactions. A field-embedded factorization machine model \cite{pande2020field} is a simple model which benefits from learning a field pair matrix embedding and higher order interactions using a DNN. 

\textbf{Attention-based models.} More recent works in CTR prediction focus on employing the multi-head self-attention technique such as automatic feature interaction learning via self-attentive neural networks (Autoint) \cite{song2019autoint}, and interpretable click-through rate prediction through hierarchical attention (InterHAt) \cite{li2020interpretable} model, which uses a multi-head self-attention block followed by hierarchical attention layers to capture higher order feature interactions. Another group of models that considers the temporal dynamics in the user behaviour, such as deep interest network (DIN) \cite{zhou2018deep} uses an attention mechanism to capture how the user interest varies towards certain advertisement and obtain adaptive representation, deep interest evolution network (DIEN) \cite{zhou2019deep} introduces the use of an attentional updated gated recurrent neural network (AUGRU) to overcome the effect of interest drifting and emphasis on relative interests. A more recent work, particularly, deep interest with hierarchical attention network (DHAN) \cite{xu2020deep} employs a hierarchical attention mechanism to model the user interest.\\
To this end, our proposed model aims to benefit from combining both multi-head self-attention and DNN components while extending their expressiveness capacity through the utilization of ReZero weighted connects for superior representation learning. Highlighting the main differences between our proposed method and the closely related work is further explained in Section 3.7.

\begin{figure*}[ht]
  \centering
  \includegraphics[scale=0.6]{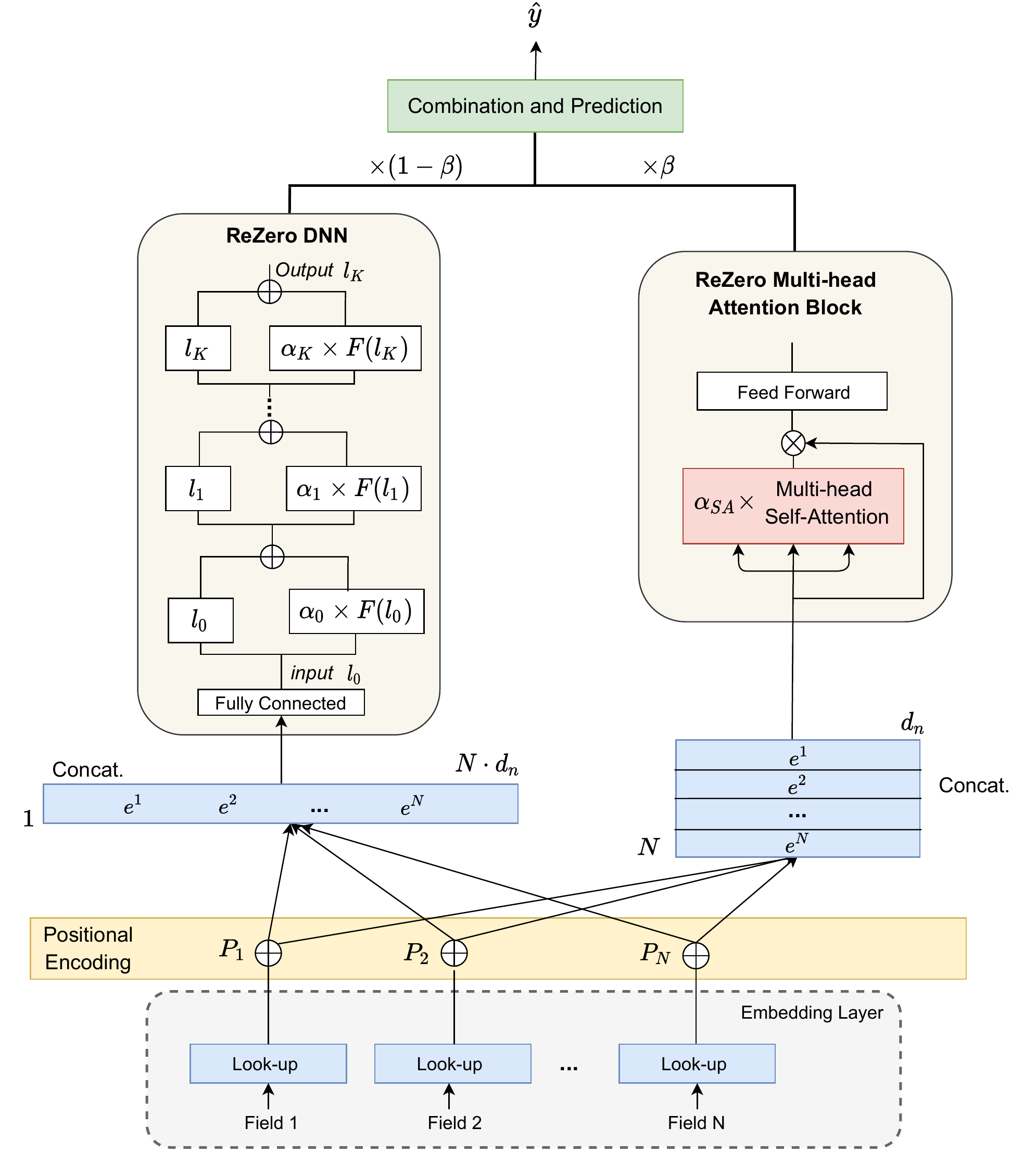}
  \caption{Deep Multi-Representation Model Architecture}
  \label{fig:1}
\end{figure*}
\section{Methodology}
\subsection{Problem Formulation}
We formulate the click-through rate prediction problem as follows;
Given a dataset of $M$ user-item interaction instances (records) $(\mathcal{X}, \mathcal{Y})$, where each instance $x_i$ in $\mathcal{X}$ has $N$ fields representing the features of the target user, the target item and the interaction's context. Moreover, each instance has a corresponding target $y_i \in \{0,1\}$, which represents if the item has been clicked by a user or not. Our main aim in the CTR task is to predict the probability of an item being clicked by a user given the input feature vector $x_i$.
\subsection{Overview}
The main goal of our proposed approach is to benefit from combining different field
representations of each input instance. As beforehand seen in the literature, stacking fully-connected layers, \cite{wang2021dcn,guo2017deepfm,huang2019fibinet} to obtain instance representations has achieved superior performance with low model complexity. Furthermore, we have seen many models now using multi-head attention to capture the relations and similarities between different fields in each instance \cite{lu2020dual,song2019autoint}. These models were able to attain remarkable performances by understanding these attention-based (multiplicative) relations. Therefore, in this work, we propose a multi-representation parallel structured model (DeepMR) for CTR prediction. The model consists of three main building blocks field embeddings, ReZero DNN, and ReZero multi-head self-attention block as shown in Figure \ref{fig:1}. Combining these components allows the model to benefit from each component's strength and reach a remarkable performance. We explain the details of each block in the forthcoming sections.
\subsection{Field Embedding Component}
To obtain the field embedding for each input instance, we use a look-up embedding table to encode the input instance fields $(x_{i}^{(1)}, x_{i}^{(2)},...x_{i}^{(N)})$. In other words, we pass all the fields through a fully connected layer to learn the latent embedding for each input field. Subsequently, we add a positional encoding to all the input fields to distinguish between the different input fields of an instance. The fields embedding can then be defined as:
\begin{equation}
    e_{i}^{(n)}= x_{i}^{(n)}W^{(n)}+P_{i}^{(n)} , \;  e_{i}^{(n)} \in \mathbb{R}^{d_n}
\end{equation}
where $W^{(n)} \in \mathbb{R}^{Z_n \times d_n}$ is the weight matrix, $Z_n$ is the vocabulary size of the input field $x_{i}^{(n)}$ and $d_n$ is the fields embedding size, where we used same $d_n$ size for all fields. $x_{i}^{(n)}$ is the binary input vector of the $nth$ field of instance $i$ and $P_{i}^{(n)}$ is the corresponding positional encoding of size $d_n$ for each field.

\subsection{ReZero Deep Neural Network Component}
Despite its simplicity, fully connected layers have shown a high capability to achieve a beneficial, meaningful representation that helps the models accomplish outstanding performances. In the ReZero-DNN component, we used a series of fully connected layers with ReZero residual connections to learn a refined representation for each input instance. Firstly, the field embeddings are concatenated column-wise to form one input vector $E_{DNN_{i}} \in \mathbb{R}^{ 1 \times ( N \cdot d_n )} $.

\begin{equation}
    E_{DNN_{i}}= \text{concat}_{col} \left( e_{i}^{(n)} \right)_{n=1:N}
    \label{eq:1}
\end{equation}

 Finally this input vector is then fed through the ReZero-DNN layers to get the final representation $O_{DNN_{i}} \in \mathbb{R}^{ d_o }$.
 \begin{equation}
    O_{DNN_{i}}= \text{ReZero-DNN}(E_{DNN_{i}})
     \label{eq:2}
\end{equation}

\textbf{ReZero Connections} Residual connections are compelling in deep learning; they were initially used in image processing with Deep ResNets \cite{he2016deep}. they ascertained performance improvement as it allows signal propagation through the network, overcome the exploding weight updates, and achieves faster convergence on the logloss. In this work, we use the novel technique residual with zero initialization \cite{bachlechner2021rezero}, which applies a simple change to the deep residual networks. Where a learnable residual weight $\alpha$ rescales the contribution of the current layer with respect to the previous one. Hence, the output of the current ReZero layer $l_{j+1}$ is:
\begin{equation}
    l_{j+1}= l_j + \alpha_j l_{j+1}
     \label{eq:3}
\end{equation}
where $\alpha_j$ is the learnable residual weight, and $l_j$ is the output of the previous layer. The empirical interpretation is further discussed in section 4.3.

\subsection{ReZero Multi-head Self-Attention Component}
Multi-head self-attention is currently used in many areas, e.g., time series forecasting, language processing \cite{vaswani2017attention}, and recommender systems \cite{kang2018self}. DeepMR also utilizes a multi-head self-attention component to enrich the extracted field representation by capturing the fields' multiplicative relations in each instance, allowing the model to learn superior representations. In this case, we concatenate the input fields embedding row-wise as shown in Figure \ref{fig:1} to obtain the input matrix $E_{SA_{i}} \in \mathbb{R}^{  N \times d_n } $ for the multi-head attention block.

\begin{equation}
    E_{SA_{i}}= \text{concat}_{row} \left( e_{i}^{(n)} \right)_{n=1:N}
     \label{eq:4}
\end{equation}

Afterward, we split the input into Query $(Q)$, Key $(K)$ and Value $(V)$ among the specified number of heads to apply the multi-head attention layer as follows: 
\begin{equation} 
\textrm{Att}(\textbf{Q}, \textbf{K}, \textbf{V}) = \textrm{softmax}\left( \frac{\textbf{Q} \textbf{K}^{T}}{\sqrt{\frac{d_o}{H}}}\right) \textbf{V}
\end{equation} 

\begin{equation} 
\begin{split}
&\textrm{SA}_i(E_{SA_{i}})= \\& 
\text{concat}_{col} \left( \textrm{Att}(E_{SA_{i}}\textbf{W}^{Q}_{h}, E_{SA_{i}}\textbf{W}^{K}_{h}, E_{SA_{i}}\textbf{W}^{V}_{h}) \right)_{h=1:H}
\end{split}
 \label{eq:5}
\end{equation} 
where $\textbf{W}^{Q}_{h}$, $\textbf{W}^{K}_{h}$, $\textbf{W}^{V}_{h} \in \mathbb{R}^{d_n \times \frac{d_o}{H}}$ represent the linear projection matrices of the head at index $h$, and $H$ is the number of heads.
$\textrm{SA}_i(E_{SA_{i}})$ represents the column-wise concatenation of the attention heads.
Finally, we have the feed-forward layers to obtain the component's final output representation $O_{SA_{i}}  \in \mathbb{R}^{ d_o }$ as follows:
\begin{equation} 
\begin{split}
&O_{SA_{i}} = \textrm{FFN}(SA_i) = \\&
\text{concat}_{row} \left( \phi (SA^{(n)}_i \textbf{W}^{(1)} + b^{(1)})\textbf{W}^{(2)} + b^{(2)}  \right)_{n=1:N}
\end{split}
 \label{eq:6}
\end{equation} 
\noindent where $\textbf{W}^{(1)}$, $\textbf{W}^{(2)} \in \mathbb{R}^{d_o \times d_o}$ are the weight matrices of the two feed-forward layers, and $b^{(1)}$, $b^{(2)} \in \mathbb{R}^{d_o}$ are their bias vectors. $\phi$ represent \textrm{Leaky\_ReLU} non-linear activation, and $\text{concat}_{row}$ concatenates vectors row-wise.

In contrast to the original multi-head self-attention additive residual connection, we apply a multiplicative ReZero connection with learnable parameter $\alpha_{SA}$ within the multi-head self-attention block, enabling the network to learn better representations. 
\subsection{Combination and Prediction}
To form the final input representation we combine the output representations from the ReZero-DNN and ReZero Multi-Head attention components using a weighted point-wise sum as follows:
\begin{equation}
    O_{{comb}_i}= \beta O_{SA_{i}} + (1-\beta)O_{DNN_{i}},\:  O_{{comb}_i} \in \mathbb{R}^{d_o}
     \label{eq:7}
\end{equation}

where $O_{DNN_{i}}$ is the output of the fully-connected ReZero-DNN layers, $O_{SA_{i}}$ is the output of the multi-head self-attention component, and $\beta$ is the weighting factor which decides the contribution of each branch to the final output representation.

Finally, the model prediction of the probability of the item $i$ being clicked by user $u$ is calculated by reducing and summing the values of the $O_{{comb}_i}$ vector elements then applying a sigmoid activation as follows:

\begin{equation}
    \hat{y}_i= \sigma(\sum_{r=1}^{d_o} O_{{comb}_i}^{(r)})
     \label{eq:8}
\end{equation}

Hence our objective function for CTR binary classification can be defined as follows;
\begin{equation}
      \mathcal{L}=  -\frac{1}{M} \sum_{i=1}^{M} y_{i} \log{(\hat{y}_{i})} + (1-y_{i}) (1-\log{(\hat{y}_{i})})
       \label{eq:9}
\end{equation}
where $y_i$ are the actual labels and $\hat{y_i}$ are the predicted probabilities, and M is the number of input instances.

The pseudo-code of DeepMR is described in Algorithm 1.
\begin{algorithm}
  \SetAlgoLined
    \SetKwInOut{Input}{input}
    \SetKwInOut{Output}{output}
    \SetKwComment{Comment}{$\triangleright$\ }{}
  \Input{A set of user-item interaction records $(\mathcal{X}, \mathcal{Y})$ with $N$ fields  }
  \Output{The predicted items click probabilities $\hat{\mathcal{Y}}$}
  Initialize the model parameters \\
  \For{E \text{epochs} }{
  
    Draw a list-wise batch $b$ from  $(\mathcal{X}, \mathcal{Y})$ \\
    Calculate the field embeddings $e_{i}^{(n)}$ for each instance\\
    Concatenate the field embedding column-wise to form the ReZero-DNN input $E_{DNN}$ using Eq.(2)\\
    Concatenate the field embedding row-wise to form the multi-head self-attention input $E_{SA}$ using Eq.(5)\\
    Update the ReZero-DNN layers to obtain $O_{DNN}$\\
    Jointly update the multi-head self-attention block parameters to obtain $O_{SA}$\\
    Combine both outputs using Eq.(9)\\
    Predict items click probabilities using Eq.(10)
  }
  \caption{DeepMR $(\mathcal{X}, \mathcal{Y})$}
  \label{DeepMR}
\end{algorithm}

\subsection{Relationship to Related Work}
By removing the DNN component of the model and the ReZero connections in the multi-head self-attention block, our model will be very similar to the Autoint model \cite{song2019autoint}. In contrast, to DIFM, we do not employ re-weighting layers or the factorization machines for the final prediction layer. However, we similarly use a self-attention component and a fully connected layers component, but with weighted residual connections on the fully connected layers and multiplicative weighted residual connections in the self-attention block. Additionally, we use a different combination layer with a fixed or learnable parameter $\beta$.

\section{Experiments}
In this section, we conducted experiments to answer the following research question;\\
\textbf{RQ1} How well does the model perform against the state-of-the-art models? \\
\textbf{RQ2} What is the impact of adding the ReZero connection to the model? \\
\textbf{RQ3} What are the impacts of the different components of DeepMR? \\
\textbf{RQ4} What is the effect of learning the combinatorial factor Beta? \\


\begin{table*}[ht]
\centering
\begin{tabular}{c|cc|cc|cc}
\hline

Method    & \multicolumn{2}{c}{Frappe} & \multicolumn{2}{c}{MovieLens} & \multicolumn{2}{c}{LastFM}\\
            & AUC & Logloss & AUC & Logloss & AUC & Logloss\\
\hline  
DeepFM \cite{guo2017deepfm}     & 0.9713 $\pm$ 3e-3 &  0.1922 $\pm$ 1e-2 & 0.9399 $\pm$ 5e-5& 0.2846 $\pm$ 1e-3 & 0.9388 $\pm$ 6e-4& 0.3038 $\pm$ 2e-3 \\
NFM \cite{he2017neural}       & 0.9780 $\pm$ 1e-3  &  0.1955 $\pm$ 3e-3 & 0.9368 $\pm$ 6e-4 & 0.3378 $\pm$ 4e-3& 0.9356 $\pm$ 3e-3 & 0.3428 $\pm$ 3e-3 \\
xDeepFM \cite{lian2018xdeepfm}     & 0.9803 $\pm$ 1e-4 & 0.1795 $\pm$ 3e-3 & \underline{0.9570 $\pm$ 3e-4}  & 0.2365 $\pm$ 9e-4& 0.9469 $\pm$ 1e-3& \underline{0.2847 $\pm$ 2e-3 } \\
AutoInt+ \cite{song2019autoint}    & 0.9797 $\pm$ 4e-4  &  0.1679 $\pm$ 4e-3 & 0.9419 $\pm$ 3e-4 & 0.3023 $\pm$ 3e-3 & 0.9388 $\pm$ 2e-3 & 0.3317 $\pm$ 4e-3   \\
FiBiNet \cite{huang2019fibinet}     & 0.9781 $\pm$ 2e-3  &  0.2093 $\pm$ 4e-3 & 0.9434 $\pm$ 1e-3 & 0.2894 $\pm$ 1e-2 & 0.9398 $\pm$ 1e-3 & 0.3432 $\pm$ 3e-3  \\
DIFM \cite{lu2020dual}       & 0.9778 $\pm$ 1e-3 &  0.1814 $\pm$ 1e-3 & 0.9569 $\pm$ 3e-4 & \underline{0.2363 $\pm$2e-4}  & 0.9441 $\pm$ 7e-4& 0.2871 $\pm$ 2e-3 \\
DCN-Mix \cite{wang2021dcn}     & \underline{0.9806 $\pm$6e-4}   &  \underline{0.1673 $\pm$ 6e-4}  & 0.9536 $\pm$ 6e-4 & 0.2633 $\pm$ 4e-3  &  \underline{0.9455 $\pm$5e-3} & 0.2907 $\pm$ 1e-2  \\
\hline
\textbf{DeepMR (ours)} & \textbf{0.9851* $\pm$ 34e-5 }  & \textbf{0.1395* $\pm$ 84e-5} & \textbf{0.9654* $\pm$ 73e-5}   & \textbf{0.2293* $\pm$ 26e-4} & \textbf{0.9592* $\pm$ 16e-5} & \textbf{0.2840 $\pm$ 56e-5 }  \\
\hline
\end{tabular}
\\ \small{Significantly outperforms the best baseline at the *0.01 level}
\caption{Model performance and comparison against baselines. Bold represents the best performance, and underline represents the second-best obtained results.}
\label{tab:2}
\end{table*} 

\begin{table*}[!]
\centering
\begin{tabular}{c|cc|cc|cc}
\hline

Method    & \multicolumn{2}{c}{Frappe} & \multicolumn{2}{c}{MovieLens} & \multicolumn{2}{c}{LastFM}\\
            & AUC & Logloss & AUC & Logloss & AUC & Logloss\\
\hline  
DeepMR w/ ReZero  &  0.9853 & 0.1395 &   0.9651 & 0.2293 &  0.9590 & 0.2840   \\
DeepMR w/o ReZero & 0.9849 & 0.1657 &  0.9624 & 0.2322 & 0.9609 & 0.3141  \\    
\hline
\end{tabular}
\caption{Effect of adding ReZero connections on the model performance.}
\label{tab:3}
\end{table*}

\subsection{Experiment Settings}
\subsubsection{Datasets}
We evaluate the CTR prediction task on three public datasets: the Frappe mobile app usage dataset, which has been used for context-aware mobile-app recommendation. It includes user IDs, App IDs, and eight contextual features such as weather, city, etc. Secondly, the MovieLens dataset\footnote{\url{https://github.com/hexiangnan/neural_factorization_machine/tree/master/data}} focuses on the personalized tag recommendation; each instance includes the user ID, Movie ID, and tag. Furthermore, we study the musician listing LastFM dataset\footnote{https://grouplens.org/datasets/hetrec-2011/}, we learn the personalized user-artist tag recommendation, it contains the user ID, Artist ID, tag, and date information such as day, month, and year. We applied the same feature pre-processing as in \cite{he2017neural,yu2019input}.
\begin{table}[H]
\centering
\begin{tabular}{c|c|c|c|c}
\hline
Dataset    & Instances & User\# & Item\# & Fields \\
\hline
Frappe      & 288,609  & 957  & 4,082  &  10  \\
MovieLens    & 2,006,859 & 17,045 & 23,743 &  3  \\
LastFM     & 186,479 & 2099 & 18,744 &  6  \\
\hline
\end{tabular}
\caption{Datasets Statistics}
\label{tab:1}
\end{table}

\subsubsection{Evaluation Protocol}
We follow the same evaluation protocol as \cite{he2017neural,yu2019input} by randomly sample two negative items that the user has not interacted with for each data instance. We split each of the three datasets into 8:1:1 for training, validation, and testing, respectively. The datasets statistics are summarized in Table \ref{tab:1}. For evaluation, we adopted the Area Under the Curve \textbf{(AUC)} which indicates the score of the model's ability to assign higher scores to positive items than negative items,  and \textbf{logloss} defined in equation \ref{eq:9} which we aim to minimize. We report the mean and standard deviation of 5 runs of the best hyper-parameter setting.

\begin{table*}[!]
\centering
\begin{tabular}{c|cc|cc|cc}
\hline

Method    & \multicolumn{2}{c}{Frappe} & \multicolumn{2}{c}{MovieLens} & \multicolumn{2}{c}{LastFM}\\
            & AUC & Logloss & AUC & Logloss & AUC & Logloss\\
\hline  
DeepMR Multi-ReZero &  0.9853 & 0.1395 &   0.9651 & 0.2293 &  0.9590 & 0.2840   \\
DeepMR Add-ReZero & 0.9838 & 0.1404   & 0.9551   &  0.2699  & 0.9565 & 0.2833  \\    
\hline
\end{tabular}
\caption{Effect of using additive vs multiplicative ReZero connections on the model performance.}
\label{tab:4}
\end{table*}

\subsubsection{Baselines}
We compare the proposed method against the following state-of-the-art models for CTR prediction.\\
\begin{itemize}
\item \textbf{DeepFM} \cite{guo2017deepfm}: A model with wide and deep architecture, which learns the features interactions using factorization machines and three multi-layer perceptron (MLP) layers.  
\item \textbf{NFM} \cite{he2017neural}: Learns the second order feature interactions from FM along with non-linear neural networks interactions.  
\item \textbf{xDeepFM} \cite{lian2018xdeepfm}: An updated version of the DeepFM model that utilizes a compressed interaction network and a DNN to capture the higher order feature interactions in an explicit and implicit fashion. 
\item \textbf{FiBiNet} \cite{huang2019fibinet}: A model which utilizes squeeze-excitation network (SENET) to learn feature importance and fine-grained interactions.  
\item \textbf{Autoint+} \cite{song2019autoint}: A model which uses multi-head self-attention to capture high-order feature interactions.
\item \textbf{DIFM} \cite{lu2020dual}: A dual input-aware factorization machines model that leverage deep neural networks (DNN) along with multi-head self-attention and residual networks to obtain informative representations of the instances.
\item \textbf{DCN-Mix} \cite{wang2021dcn}: A state-of-the-art mixture model of low-rank deep and cross networks for CTR prediction.
\end{itemize}



\textbf{Hyperparameters Settings}. We ran our experiments using GPU RTX 2070 Super and CPU Xeon Gold 6230 with RAM 256 GB. We used Tensorflow\footnote{https://www.tensorflow.org} for implementation; for reproducibility purposes, our code is available here \footnote{\url{https://github.com/Shereen-Elsayed/DeepMR}}. For running all baselines, we used the publicly available DeepCTR\footnote{https://github.com/shenweichen/DeepCTR} library. For selecting the model hyperparameters, we adopted a grid search on the field embeddings between [30 - 600], on the learning rate between [0.000005 and 0.0001], the L2-regularization lambda between [0.0001 and 0.2], and the dropout rate between [0.01 and 0.9]. The best parameters for dropout, L2-regularization, embedding size and learning rate respectively are 0.35, 0.03, 512 and 9e-05 for Frappe dataset, while for LastFM dataset 0.35, 0.05, and 7e-06 with 512 embedding size, finally for MovieLens dataset 0.45, 0.009 and 8e-05 with 600 latent embedding size. We used the ReLU activation function for all the models' layers and the Leaky ReLU activation function in the feed-forward layer within the attention block, which has shown better performance. Finally, we employ Adam\cite{kingma2014adam} for optimizing the proposed model. 

\begin{table*}
\centering
\begin{tabular}{c|cc|cc|cc}
\hline

Method    & \multicolumn{2}{c}{Frappe} & \multicolumn{2}{c}{MovieLens} & \multicolumn{2}{c}{LastFM}\\
            & AUC & Logloss & AUC & Logloss & AUC & Logloss\\
\hline 
DeepMR  &  0.9853 & 0.1395 &   0.9651 & 0.2293 &  0.9590 & 0.2840   \\
DeepMR With $\beta_L$ & 0.9856   & 0.1408   &  0.9611 & 0.2260  & 0.9589 & 0.2850  \\    
\hline
\end{tabular}
\caption{Effect of learning the factor Beta on the model performance.}
\label{tab:5}
\end{table*}

\begin{figure*}
     \centering
     \begin{subfigure}[b]{0.45\textwidth}
         \centering
         \includegraphics[scale =0.4]{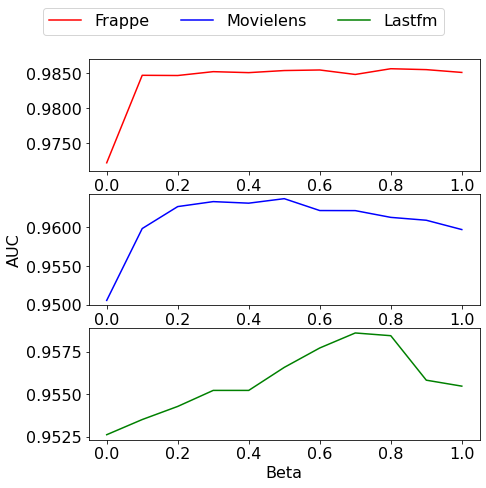}
     \end{subfigure}
     \begin{subfigure}[b]{0.45\textwidth}
         \centering
         \includegraphics[scale=0.4]{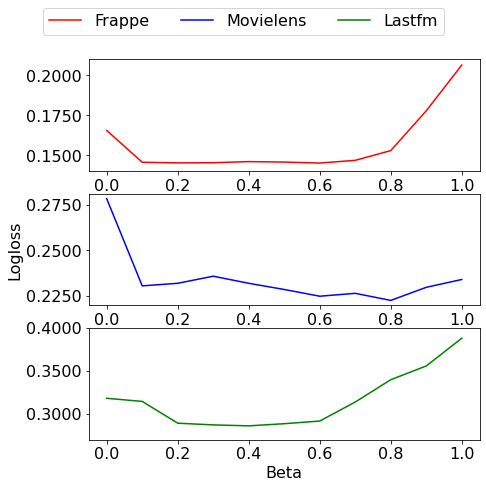}
     \end{subfigure}    
\caption{Effect of changing Beta on the model performance.}
\label{fig:2}
\end{figure*}

\subsection{Model Performance and Comparison Against Baseline (RQ1)}
In this section, we report the results of the proposed DeepMR method and compare its performance against state-of-the-art models. The results reported in Table \ref{tab:2} indicate the effectiveness of the proposed method and its ability to outperform all baselines on the AUC metric with a considerable improvement. While on the logLoss, DeepMR outperforms other models on Frappe and MovieLens datasets significantly while showing comparable results on the LastFM dataset logloss. However, other methods which incorporate different components to obtain high-order feature interactions, e.g., xDeepFM, achieve the second-best AUC performance on the MovieLens dataset. While including a Mixture of Low-rank Deep and Cross Network (DCN-Mix) has shown high competitiveness on the Frappe dataset and the second-best AUC on the LastFM dataset. The DIFM model utilizes multi-head self-attention and stacked fully connected layers, and Factorization machine components show better performance than AutoInt+ on the MovieLens dataset.
In contrast, AutoInt+, which relies on the multi-head self-attention, outperforms DeepFM and DIFM on the Frappe dataset. The performance of the CTR models highly depends on the datasets; thus, the obtained results differ from one dataset to another. Concerning the logloss, the best performances were achieved by DCN-Mix on the Frappe dataset, DIFM on the MovieLens dataset, and xDeepFM on the LastFM dataset. 


\begin{figure*}
     \centering
     \begin{subfigure}[b]{0.45\textwidth}
         \centering
         \includegraphics[scale=0.4]{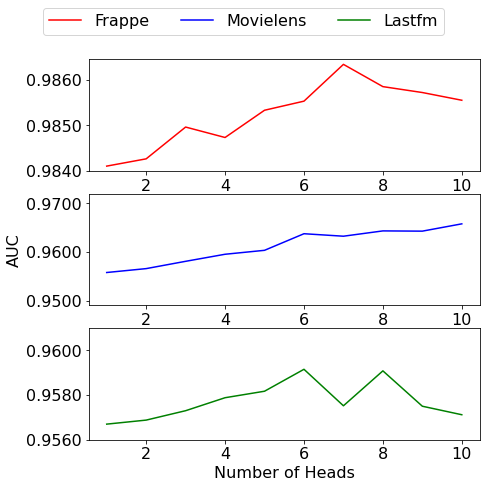}
     \end{subfigure}
     \begin{subfigure}[b]{0.45\textwidth}
         \centering
         \includegraphics[scale=0.4]{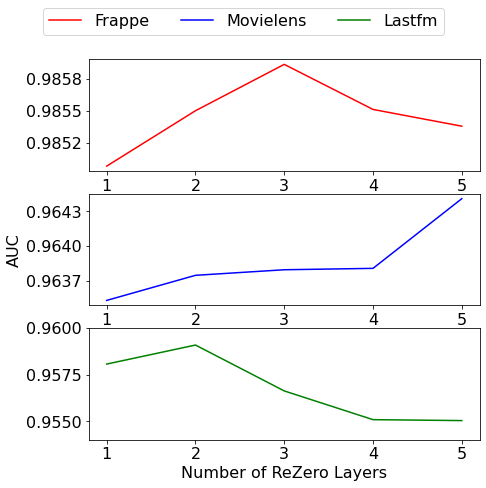}
     \end{subfigure}
\caption{Effect of different Beta values, number of layers and number of heads on the AUC.}
\label{fig:3}
\end{figure*}
\begin{figure*}
     \centering
     \begin{subfigure}[b]{0.45\textwidth}
         \centering
         \includegraphics[scale=0.4]{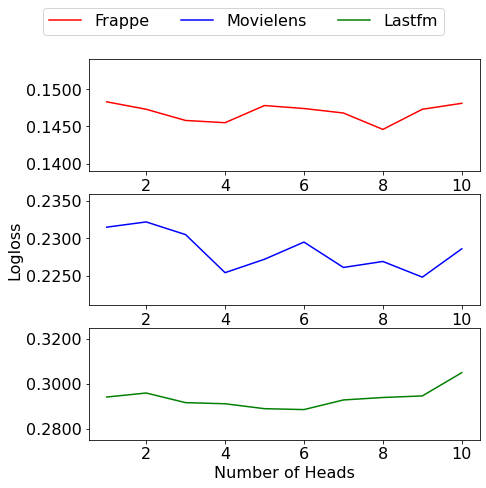}
     \end{subfigure}
     \begin{subfigure}[b]{0.45\textwidth}
         \centering
         \includegraphics[scale=0.4]{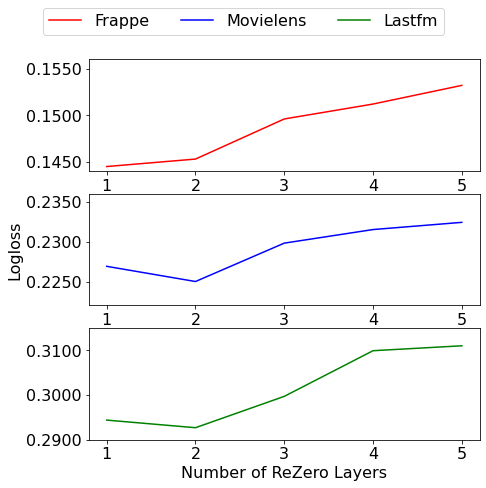}
     \end{subfigure}
\caption{Effect of different Beta values, number of layers and number of heads on the Logloss.}
\label{fig:4}
\end{figure*}

\begin{figure}
     \centering
         \includegraphics[scale =0.4]{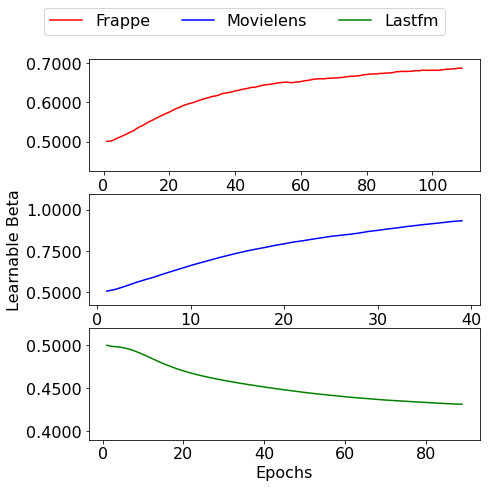}
         \caption{Effect of learning Beta over epochs.}
         \label{fig:5}
\end{figure}

\subsection{Effect of Adding ReZero Connections (RQ2)}

In this section, we study the effect of adding the weighted residual connections with learnable weights to the DNN and the multi-head self-attention blocks. First, we removed all ReZero connections in the DNN and the attention block, and we kept the residual connection that existed in the original transformer architecture \cite{vaswani2017attention}. Results in Table \ref{tab:3} show that the residual connections significantly improved the model's performance across the Frappe and MovieLens datasets. On the other hand, it was not very effective on the LastFM dataset. Regarding the logloss metric, it is clear that the ReZero connections affected the logloss performance significantly, which is one of the main benefits of applying ReZero connections which help improve performance and faster convergence.\\
Furthermore, we investigated the effect of using the multiplicative residual connection within the multi-head self-attention block versus additive connections. Multiplicative residuals were applied in various image processing operations \cite{zhang2018image} and positively impacted the obtained results. Results in Table \ref{tab:4} show that adding multiplicative ReZero connections within the attention block positively impacted the model performance on all datasets on the AUC. Similarly, the logloss fell behind when using the additive residuals except on the LastFM dataset, which has shown slight improvement from 0.2840 to 0.2833. Finally, it is worth mentioning that we applied the multiplicative connections on the DNN component, but it was not as effective in that case.

\subsection{Impact of Combinatorial Factor Beta (RQ3)}

Given that the proposed model combines two components, particularly, it can be considered a mixture of experts with setting the number of experts to two. We combine the final output with the weighted sum of the two model branches. Therefore, each model part contributes to the final output probabilities. Figure \ref{fig:2} illustrates the effect of changing $\beta$, which varies from $0$ to $1$, where $\beta =0$ means that the ReZero-DNN is the active branch while $\beta = 1$ means that the multi-head self-attention is the functional one. As shown in Figure \ref{fig:2} the best $\beta$ value differs from one dataset to another as for Frappe, and LastFM datasets best values are 0.8 and 0.7, while for MovieLens, the best value is at 0.5. However, for most of the datasets, it is clear that the multi-head self-attention branch has a higher contribution to the final output, specifically for Frappe and MovieLens datasets. 

Besides, Figure \ref{fig:2} shows the influence of varying the combinatorial weight on the Logloss. For the Frappe dataset, the differences in the Logloss are minimal between the rate 0.2 and 0.6. In contrast, the best Logloss is acquired at 0.6, concerning the MovieLens dataset, the model attained the best Logloss at 0.8, finally for the LastFM dataset selecting a less weight 0.4 had the best Logloss.

\subsection{Impact of Learning Combinatorial Factor Beta (RQ4)}

As explained earlier, our model is a mixture of two components, each of those two branches has a certain impact on the final output of the model. Such that the final output is the weighted sum of the two components as shown in Equation \ref{eq:7}, where the weight ($\beta$) adjusts the contribution of each component. We have shown in the previous section an ablation study of the impact of changing the weights and how it can affect the final performance consequently. Another way of adapting the contribution weight of each component is to learn how each part impacts the model performance. In other words, we can learn the weighting factor $\beta$ along with the models' parameters to optimize our objective function. To this end, we initialize the weight $\beta$ with 0.5 and set it to be trainable simultaneously with the model parameters. Table \ref{tab:5} illustrate the obtained results of learning the combinatorial factor $\beta$ with we denote as $\beta_L$ . As shown in Table \ref{tab:5} the results, there is a positive impact of learning the weight on the Frappe dataset; alternatively, for the MovieLens and LastFM datasets, it rather caused a reduction in model AUC. Figure \ref{fig:5} show how learning $\beta$ changes over epochs; for Frappe and MovieLens datasets, beta steadily increased to 0.66 for Frappe and 0.92 for MovieLens, while for the LastFM dataset, it slightly decreased from 0.5 to 0.44.




\subsection{Model Settings}

In this section, we study different model settings and the effect of each parameter on the model performance. Firstly, for the multiplicative ReZero-DNN component, we study the effect of having different number of ReZero layers in the model. As shown in Figure \ref{fig:3} including three layers after the embedding layer for Frappe and two for LastFM datasets are sufficient to achieve the best performance. In contrast, the MovieLens dataset's five layers showed the best performance. Regarding the multi-head self-attention block, we report the effect of using different self-attention heads in Figure \ref{fig:3} which exemplify the impact of changing the number of heads. The best number of heads varies from one dataset to another for the LastFM dataset; six heads obtained the best result. For the Frappe dataset, the performance increased when adding more heads, reaching the best result at seven heads. Finally, the MovieLens dataset required ten self-attention heads.

As aforementioned, the model performance is highly dependent on the dataset and the optimized metric. Figure \ref{fig:4} illustrates different logloss performances based on the multi-head self-attention block's different number of heads and the number of ReZero layers in the DNN. The model achieves the best logloss performance by using a few layers, as the ReZero connections also allow the model to converge faster. The best number of layers were 1, 2, 2 for Frappe, MovieLens and LastFM, respectively. Concerning the number of self-attention heads employed, the best number varies between the datasets. However, we can notice that 8, 9, and 6 heads were the best-selected values for Frappe, MovieLens and LastFM, respectively.

\section{Conclusion}
This paper proposes DeepMR, a mixture of experts model that integrates the benefits of two feature representation components: a ReZero DNN and a multi-head self-attention component. We leverage the non-linearity in the DNN to capture the additive high-order feature interactions. We incorporate weighted residual connections (ReZero) to the DNN, enabling the model to propagate the initial signal through the layers, achieve better field representations, and enhance model performance. On the other hand, we capture high-order multiplicative feature interactions using the multi-head self-attention component. Ultimately, we conducted experiments on three publicly available datasets; results show that the proposed model outperforms state-of-the-art models consistently.

\bibliographystyle{ACM-Reference-Format}
\bibliography{sample-base}

\end{document}